\documentclass[fleqn,times]{article}
\usepackage{graphics,epsfig}
\usepackage{nips01}
\usepackage{graphicx}
\usepackage{epstopdf}
\newcommand{\beq}[1]{\begin{equation}\label{#1}}
\newcommand{\eeq}{\end{equation}}
\newcommand{\Var}[1]{{\rm Var}\left[ #1 \right]}

\newfont{\msym}{msbm10}

\def\beqa{\begin{eqnarray}}
\def\eeqa{\end{eqnarray}}
\newcommand{\ignore}[1]{} 

\newcommand{\eqref}[1]{Eq.~(\ref{#1})}

\newcommand{\figref}[1]{Fig.~\ref{#1}}
\newcommand{\tabref}[1]{Table~\ref{#1}}
\newcommand{\secref}[1]{Section~\ref{#1}}

\begin{document}

\title{Estimating mutual information and multi--information in large networks}

\author{Noam Slonim, Gurinder S. Atwal, Ga\v{s}per Tka\v{c}ik, and William Bialek\\
Joseph Henry Laboratories of Physics, and\\
Lewis--Sigler Institute for Integrative Genomics\\
Princeton University, Princeton, New Jersey 08544\\
\{nslonim,gatwal,gtkacik,wbialek\}@princeton.edu}

\maketitle

\begin{abstract} 
We address the practical problems of estimating the  information relations that characterize large networks.  
Building on  methods developed for  analysis of the neural code, 
we show that reliable estimates of mutual information can be obtained with manageable computational effort.
The same methods allow estimation of higher order, multi--information terms.
These ideas are illustrated by analyses of gene expression, financial markets, and consumer preferences.  In each case, information theoretic measures correlate with independent, intuitive measures of the underlying structures in the system.
\end{abstract}

\section{Introduction}

Many problems of current scientific interest are described, at least colloquially, as being about the flow and control of information.  It has been more than fifty years since Shannon formalized our intuitive notions of information, yet relatively few fields actually use information theoretic methods for the analysis of experimental data.  Part of the problem is practical:  Information theoretic quantities are notoriously difficult to estimate from limited samples of data, and the problem expands combinatorially as we look at the relationships among more variables.  Faced with these difficulties most investigators resort to simpler statistical measures (e.g., a correlation coefficient instead of the mutual information), even though the choice of any one such measure can be somewhat arbitrary.  Here we build on methods developed for the information theoretic analysis of the neural code, and show how the practical problems can be tamed even in large networks where we need to estimate millions of information relations in order to give a complete characterization of the system.  To emphasize the generality of the issues, we give examples from analyses of gene expression, financial markets, and consumer preferences.

In the analysis of neural coding we are interested in estimating (for example) the mutual information between sensory stimuli and neural responses.  The central difficulty is that the sets of possible stimuli and possible responses both are very large, and sampling the joint distribution therefore is difficult.  Naively identifying observed frequencies of events with probabilities leads to systematic errors, underestimating entropies and overestimating mutual information.  There is a large literature about how to correct these errors, going back to Miller's calculation of their magnitude in the asymptotic limit of large but finite sample size \cite{miller}.  Strong et al \cite{Strong+al98} showed that the mutual information between stimuli and responses could be estimated reliably by making use of two ideas.  First, averages over the distribution of stimuli were replaced with averages over time, using ergodicity.  Second, the sample size dependence of information estimates was examined explicitly, to verify that the data are in the asymptotic limit and hence that one can extrapolate to infinite sample size as in Miller's calculation;   extrapolating each data set empirically, rather than applying a universal correction, avoids assumptions about the independence of the samples and the number of  responses that occur with nonzero probability.   This has come to be called the ``direct method'' of information estimation.

One of the central questions in neural coding is whether the precise timing of action potentials carries useful information, and so it makes sense to quantize the neural response at some fixed time resolution and study how the mutual information between stimulus and response varies as a function of this resolution.  In other contexts quantization is just a  convenience (e.g., in estimating the mutual information among gene expression levels), but there is an interaction between the precision of our quantization and the sample size dependence of the information.  An additional challenge is that we want to estimate not the mutual information between the stimulus and the response of one individual neuron, but  the information relations among the expression levels of thousands of different genes; for these large network problems we need more automated  methods of insuring that we handle correctly all of the finite sample size corrections.  Finally, we are interested in more than pairwise relations; this poses further challenges that we address here.

\section{Correcting for finite sample size}
\label{Sec:Direct}

Consider two vectors, 
$\vec{y}_i = \{ y_i(1), y_i(2), \cdots , y_i(N)\}$ and
$\vec{y}_j = \{ y_j(1), y_j(2), \cdots , y_j(N)\}$, that represent the expression levels of two genes  in $N$ conditions.  We view these observations as having been drawn out of the joint probability density $P_{ij}(y_i , y_j)$ of expression levels that the cell generates over its lifetime or (in practice) over the course of an experiment.
Information theory tells us that there exists a unique 
measure of the interdependence of the  two expression levels,
the Mutual Information (MI):
\begin{equation}
\label{Eq:MI}
I[P_{ij}] = \int d y_i \int d y_j \;P_{ij}(y_i,y_j) \log_2 \left[\frac{P_{ij}(y_i,y_j)}{P_i(y_i)P_j(y_j)}\right]~,
\end{equation}
where $P_i$ and $P_j$ are  the marginal distributions.
Recall that  $I[P_{ij}]$ quantifies how much information  the expression level of one gene provides about the expression level of the other, and   is invariant to any invertible transformation of the individual variables.

\begin{figure}[] 
\begin{center}
\begin{tabular}{cc} 
    \psfig{figure=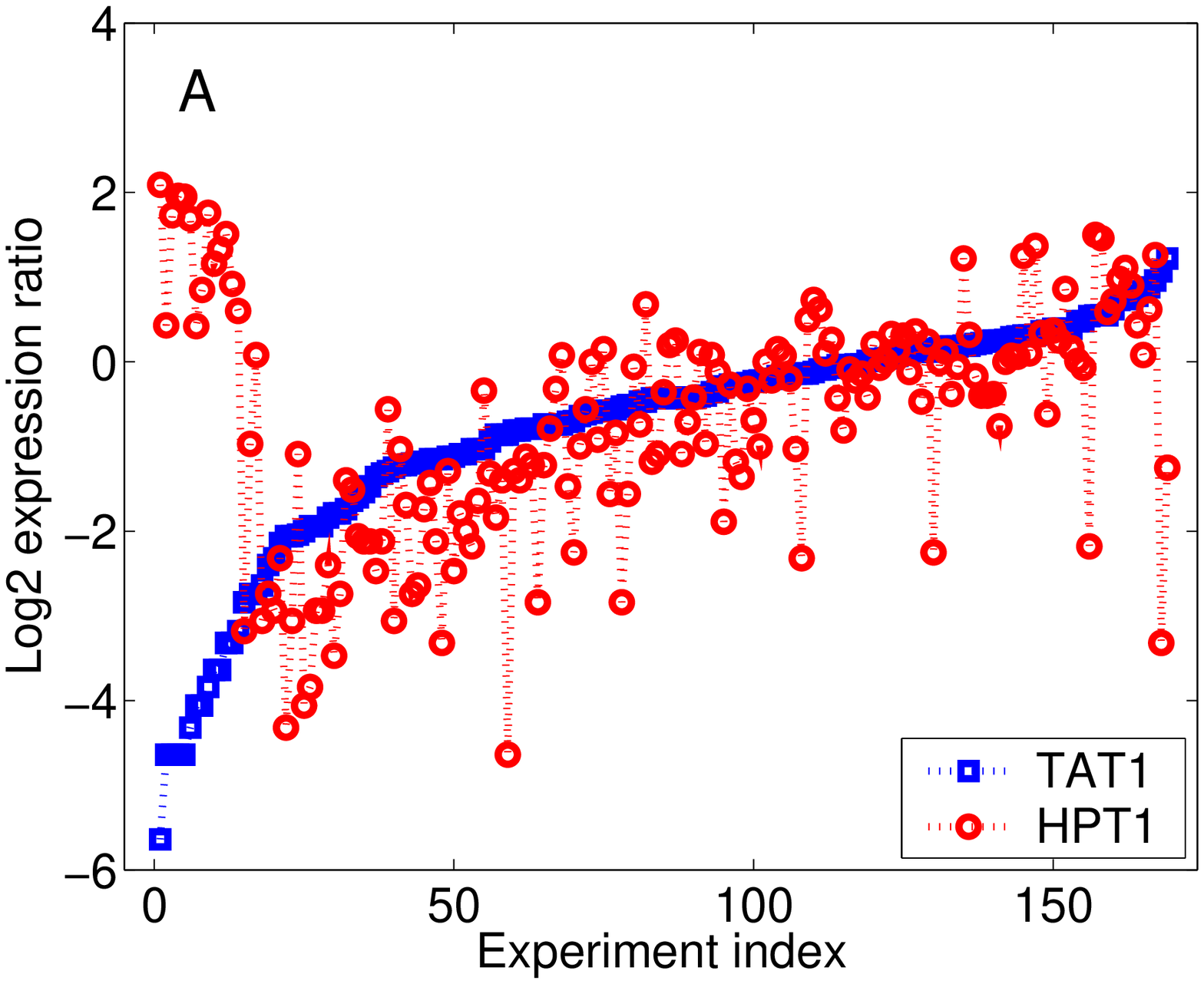,width=.45\columnwidth,height=1.5in} &
    \psfig{figure=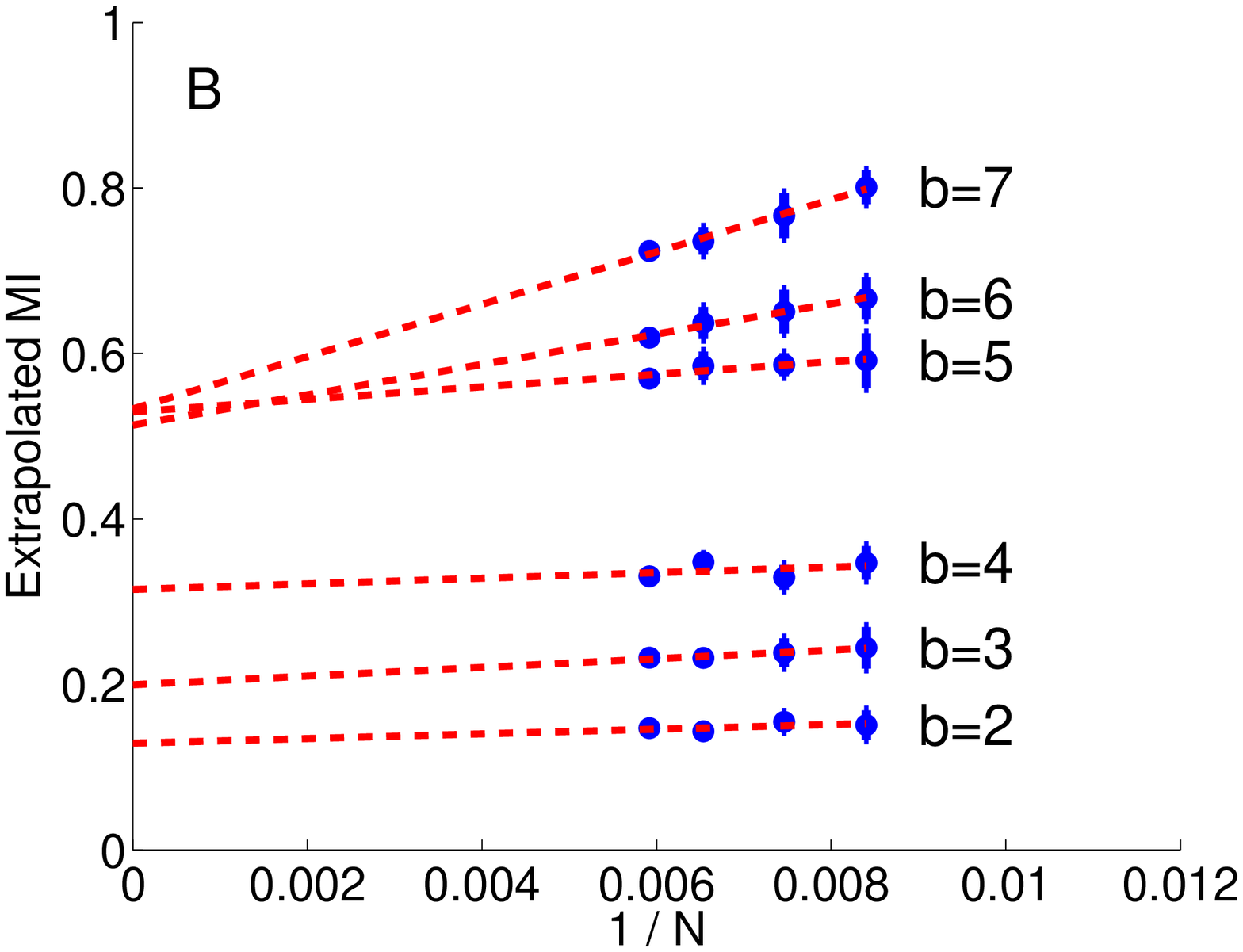,width=.45\columnwidth,height=1.5in} \\
    \psfig{figure=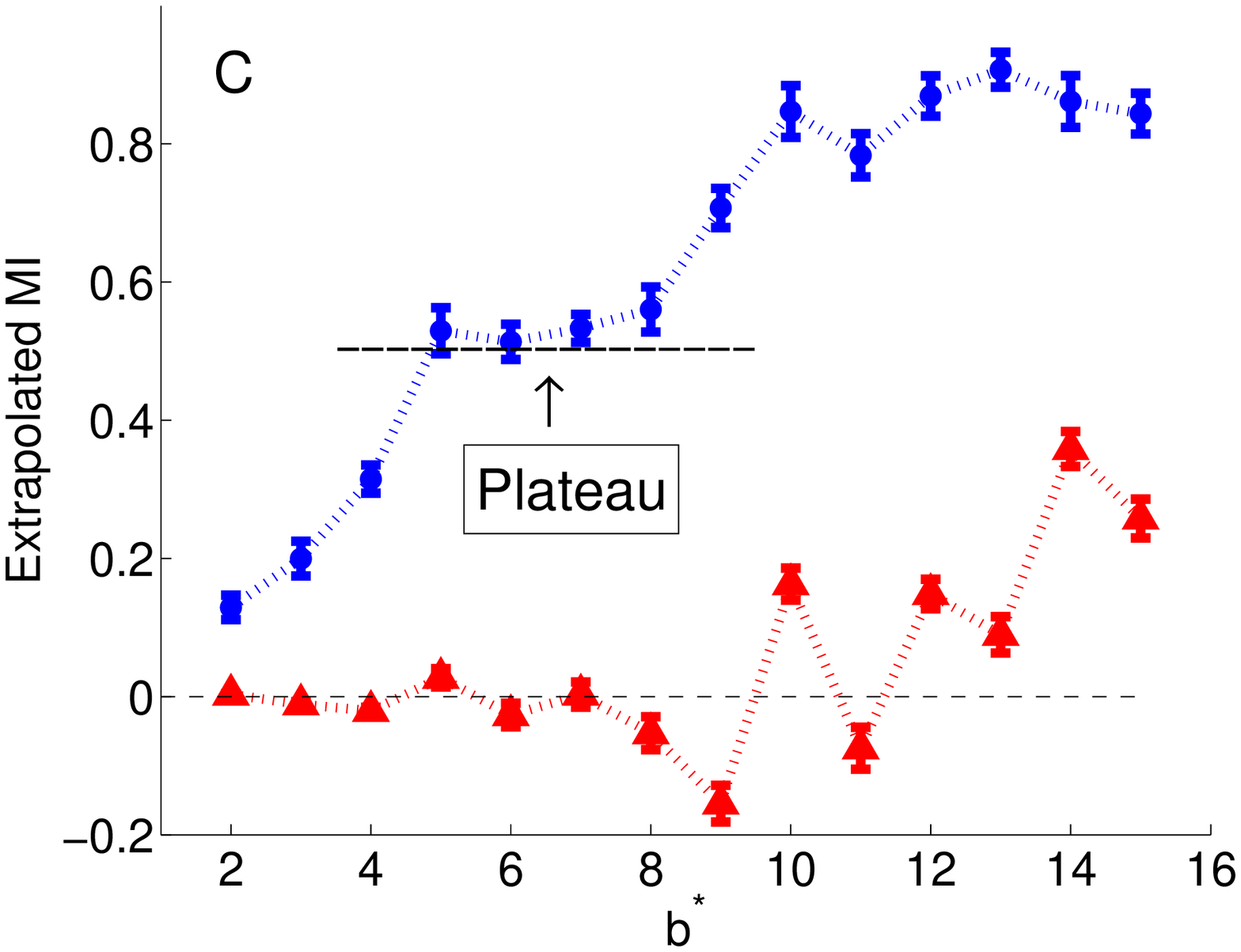,width=.45\columnwidth,height=1.5in} &
    \psfig{figure=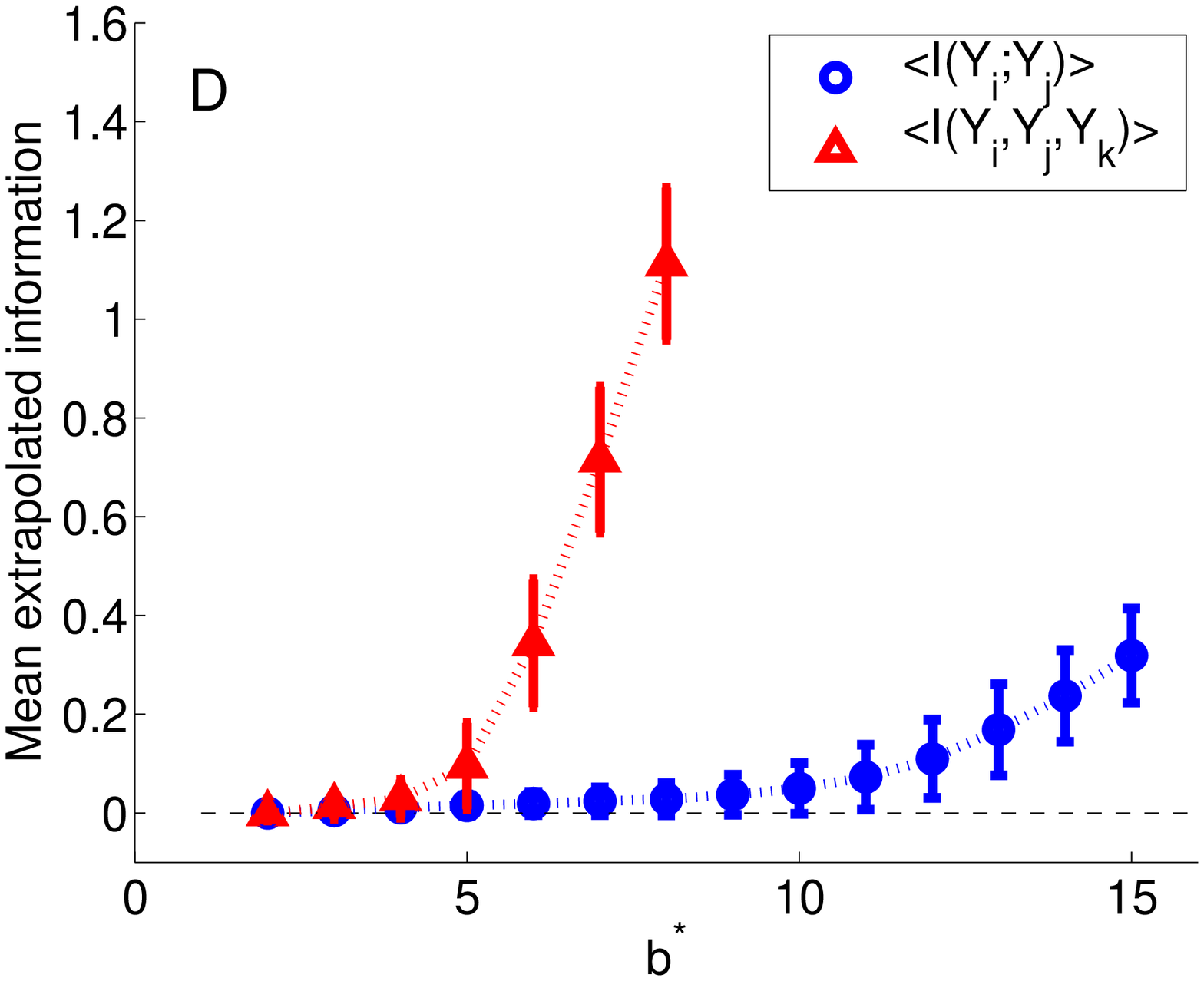,width=.45\columnwidth,height=1.5in} 
\end{tabular}
\end{center}
\caption{\small
{\bf (A)} Original expression profiles of two yeast genes,  TAT1  
and HPT1,  
responding to different stress conditions.
Data taken from \cite{Gasch00} and sorted with respect to TAT1 profile. 
{\bf (B)} Different linear extrapolation curves (for different 
quantization levels $b$) estimate the MI
between the two expression profiles.  
The extrapolated MI is $\approx 0.5\,{\rm bits}$ while the
Pearson Correlation is only $0.05$. 
{\bf (C)} Extrapolated MI values for the same pair as a function
of the  quantization level, $b$. 
The lower curve serves as a reference in which the same estimation
procedure is applied but the expression profiles are randomly reshuffled.
For $b \geq 9$ the results seem to represent overestimates.
Error bars 
represent standard deviation of MI estimates in the smallest sub--sample used.
{\bf (D)} Average estimated pairwise/triplet information values for 
$10^4$ pairs/triplets of randomly reshuffled 
gene expression profiles \cite{Gasch00} as a function of  $b$.}
\label{Fig:EstI}
\end{figure}

Estimating $I[P_{ij}]$ from a finite sample requires
regularization of $P_{ij}(y_i,y_j)$; 
the simplest regularization is to make $b$ discrete bins along each axis.   If the bins have fixed size then we break the coordinate invariance of the mutual information, but if we make an adaptive quantization so that the bins are equally populated then this invariance is preserved.   From the data processing inequality \cite{Cover91} we know that the mutual information among the discrete variables must  be less than or equal to the true mutual information.  At fixed $b$,  we use the same ideas as in \cite{Strong+al98}:  the naively estimated information  will have a dependence on the sample size $N$, 
\beq{}
I_{\rm est} (b,N) = I_{ \infty}(b) + A(b)/N + \cdots~,
\eeq{}
where  $I_{\rm \infty}(b)$ is our extrapolation to infinite sample size.  Finite size effects are larger when the space of responses is larger, hence $A(b)$ increases with $b$, and beyond some critical value $b^*$ the terms $\cdots$ become important and we lose control of the extrapolation $N \rightarrow \infty$.   We define $b^*$ by analyzing data that have been shuffled to destroy any mutual information:   
$I_{\infty}^{\rm  shuffle} (b)$ is zero within error bars for $b < b^*$, but not for $b > b^*$; $I_\infty(b)$ increases, and ideally  saturates at some $b < b^*$. For examples see  \figref{Fig:EstI} B  \& C.

\section{Dealing with a large number of pairs}
\label{sec:ManyPairs}
With a small number of pairs one can assign a significant computational
effort to each $I[P_{ij}]$. From the total of $N$ samples  we can choose many different sub--samples, and we can look manually 
for the plateau in $I_{\infty}(b)$.
With a large number of pairs a different approach is required.
The first issue is how to determine the sub--sample sizes.
Since $I_{\rm est}(b,N)$ is linear in $1/N$, we get the greatest statistical power by uniform sampling in $1/N$.  
Consider, for example, three sub--sample sizes\footnote{The same idea could be applied to any number of sub--sample sizes.}
$\{f_1 N,\,f_2 N,\,f_3 N\}$ where $0<f_1<f_2<f_3<1$;   to make sure
that $\{1/{f_1 N},\,1/{f_2 N},\,1/{f_3 N}\}$
are spaced uniformly we should choose 
$f_2=({2\,f_1\,f_3})/({f_1+f_3})$;  
$f_1$ and $f_3$ must be chosen to keep all points in the linear or asymptotic regime \cite{Strong+al98}.   The same idea can be used to ask how many independent draws of $fN$ samples we should take from the total of $N$. 
If $t(f)$ is  the number of draws with $fN$ samples, the variance of the information estimate turns out to be ${\rm var}[I_{\rm  est} (b,fN)] \propto 1/[(fN)^2t(f)]$;  achieving roughly constant error bars throughout the fitting region requires $t(f) \propto 1/f^2$.

As mentioned earlier, this procedure is valid for $b\leq b^*$
where $b^*$ depends on $N$. 
Indeed, in \figref{Fig:EstI}C we see that after the plateau at $b =5, \cdots ,8$
for $b \geq 9$ the results are less stable and information is overestimated. 
How should one determine $b^*$ in general, given that 
identifying such a plateau is not always trivial? 
A simple approach is to apply the same procedure for
different $b$ values to a large number of pairs for which 
the observations are randomly reshuffled. 
Here, positive MI values merely indicate small sample
effects, not properly corrected. 
In \figref{Fig:EstI}D we present $\langle I(y_i;y_j) \rangle$
for $10^4$ pairs of randomly reshuffled
gene expression profiles \cite{Gasch00} as a function of $b$. 
Based on this figure we chose a (conservative) value
of $b^*=5$ for all pairs. Notice, though, that this approach might
yield some under-estimation effects, especially for highly informative
pairs.  
Once $b^*$ is set the procedure is completed by
estimating $I$ for each $b \leq b^*$ and choosing
the last extrapolated value that provides a significant improvement
over less detailed quantizations.\footnote{By a significant improvement we mean an improvement
beyond the error bar. A simple scheme to define
such error bars is to use the standard deviation 
of the naive MI values obtained for the smallest sub--sample
used during the extrapolation. We note that we tried 
other alternatives with no significant effect.}

\section{Estimating more than pairwise information relations}
\label{Sec:TripletEst}

The mutual information
has a natural generalization to multiple variables,
\begin{equation}
I_r[P_{1 \cdots r}(y_1, \cdots ,y_r)] = \int d^r y\,P_{1 \cdots r}(y_1, \cdots , y_r) \log_2 \left[\frac{P_{1\cdots r}(y_1, \cdots ,y_r)}{P_1(y_1)  \cdots P_r(y_r)}\right] .
\end{equation}
This  {\em multi--information}  captures more collective properties than just pairwise relations, but in the same general information theoretic framework.
It should be clear that estimating this term  
is far more challenging since the number of parameters in the relevant joint
distribution is exponential in $r$. 
Nonetheless, here we show that  triplet information values ($r=3$)
can be estimated reliably. 
We start with a ``multi--information chain rule,'' decomposing $I_r$ 
into a  sum
of $(r-1)$ mutual information terms,
\beq{}
\label{Eq:multiIbreak}
I_r[P_{1 \cdots r}(y_1,y_2, \cdots y_r)] = \sum_{r'=2}^{r} I(y_{r'-1};y_{r'}, \cdots ,y_r)~;
\eeq 
for $r=3$ we have
$I_3[P_{ijk}] = I(y_i;y_j,y_k) + I(y_j;y_k)$.
Thus, we can directly apply our procedure to estimate these two
pairwise information terms, ending up with an estimate for the triplet information. 
Note   that in $I(y_i;y_j,y_k)$ the quantized versions of $y_j$ and $y_k$
should be combined into a single quantized variable with $b^2$ bins,
hence the relevant joint distribution now consists of $b^3$ entries. 
Thus, increasing the quantization
level with limited data is more difficult and one should expect to use
a lower $b^*$ bound in order to avoid overestimates. 
Note also that there
are $3$ different ways to estimate a triplet information term, 
by permuting $i,j$ and $k$. This provides a built in 
verification scheme in which every term is estimated through these $3$ different compositions, 
and the resulting estimates are compared to each other.

\section{Applications}
\label{Sec:Applications}

\subsection{Datasets and implementation details}

The first data we consider are the expression responses of yeast genes
to various forms of environmental stress \cite{Gasch00}.
Every gene is represented by the log--ratio of expression levels in $N=173$ conditions.\footnote{Importantly, the mutual (and multi) information are invariant to any
invertible changes of variables. Thus, the log transformation has no effect on our results.}
We concentrate on $868$ genes characterized in \cite{Gasch00}
as participating in the Environmental Stress Response (ESR) module;
$283$ of these genes have increased mRNA levels in response to stressful environments
(the ``Induced'' module), and $585$ genes display the opposite 
behavior (the ``Repressed'' module). Since the responses in each group were
claimed to be almost identical \cite{Gasch00} we expect
to find mainly strong positive and negative linear correlations in these data.
In our second example we consider 
the companies in the Standard and Poor's $500$ ($SP500$) \cite{SP500}. 
Every company was represented by its day--to--day fractional changes in stock price 
during the trading days of $2003$ ($N=273$).
As our third test case we consider the $EachMovie$ dataset, 
movie ratings provided by more than $70,000$ viewers \cite{EachMovie}.
These data are inherently quantized as only six discrete possible ratings were used. 
Hence, no quantization scheme need be applied and  
we represented each movie by its ratings from different viewers
and focused on the $500$ movies that got the maximal number of votes. 
While estimating the MI for a pair of movies, only 
viewers who voted for both movies were considered. 
Hence, the sample size for different pairs
varied by more than three orders of magnitude
(ranging from $N=11$ to $N=26,220$ joint votes),
providing an interesting test of the sensitivity of our approach with respect to this parameter. 

To demonstrate the robustness of our procedure, 
in all applications we used the same parameter configuration:
extrapolation  was based on three sub--sample sizes,
$f_1=0.7 N,\,f_2=0.7875 N,\,f_3=0.9 N$, where for each sub-sample size
we performed $t_1=21,\,t_2=16,\,t_3=12$ naive estimation trials, respectively
(see \secref{sec:ManyPairs}). 
Together with the full sample size 
we ended up with a total of $50$ trials
for a single information estimation, which represented a reasonable compromise
between estimation quality and available computational resources. 
For the ESR data we found $b^*=5$ based on \figref{Fig:EstI}D, and similarly   $b^*=5$
for the $SP500$ data. 
In this configuration, estimating the pairwise information 
between many pairs is quite feasible. For example, with
$\sim 1.25\times 10^5$ pairs in the $SP500$ data 
the overall running time 
is less than two hours in a standard work station
(Linux OS, 3GHz CPU, 1GB RAM). 

\subsection{Verification schemes}

\begin{figure}[] 
\begin{center}
\begin{tabular}{ccc} 
   \psfig{figure=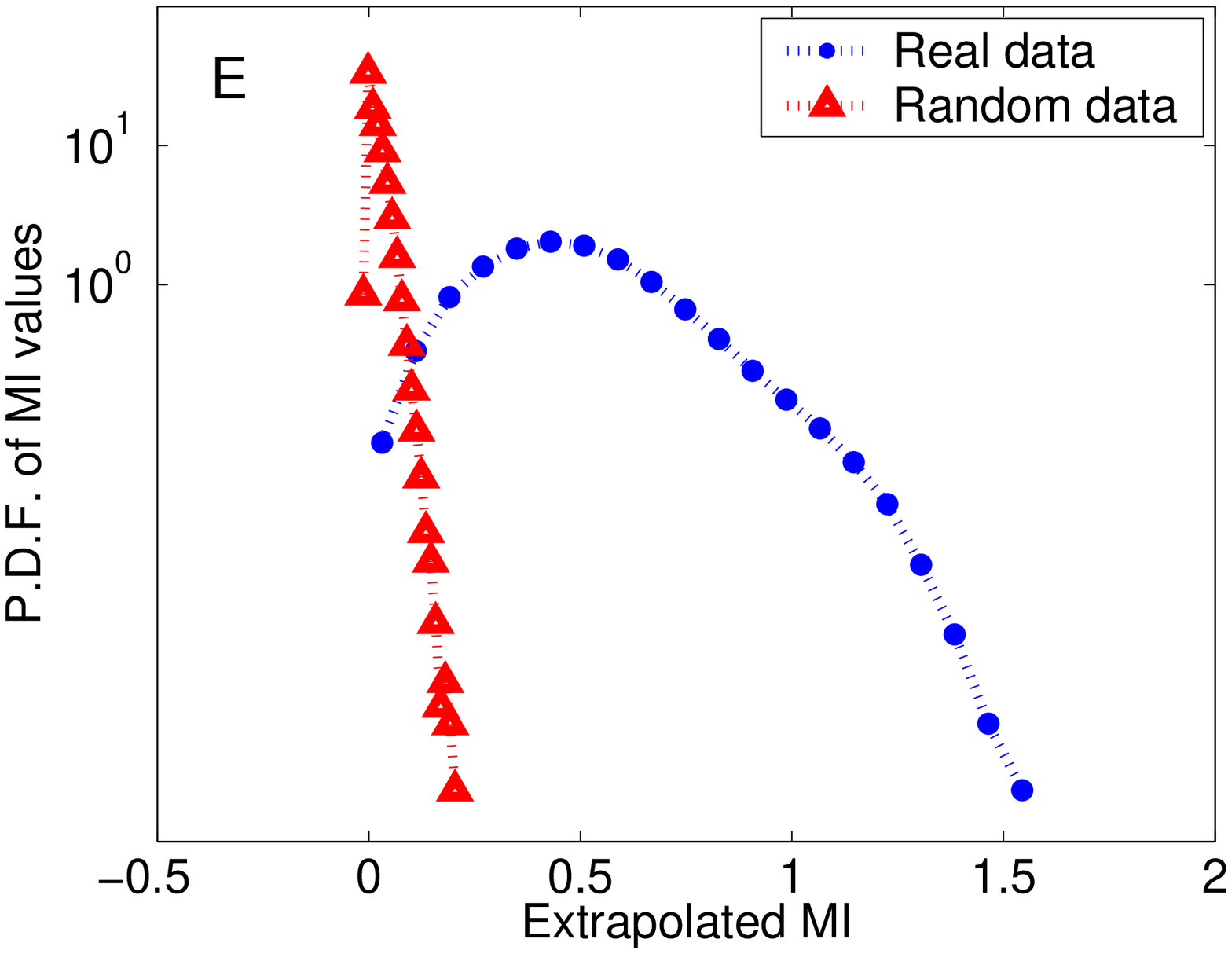,width=.45\columnwidth,height=1.5in} &
   \psfig{figure=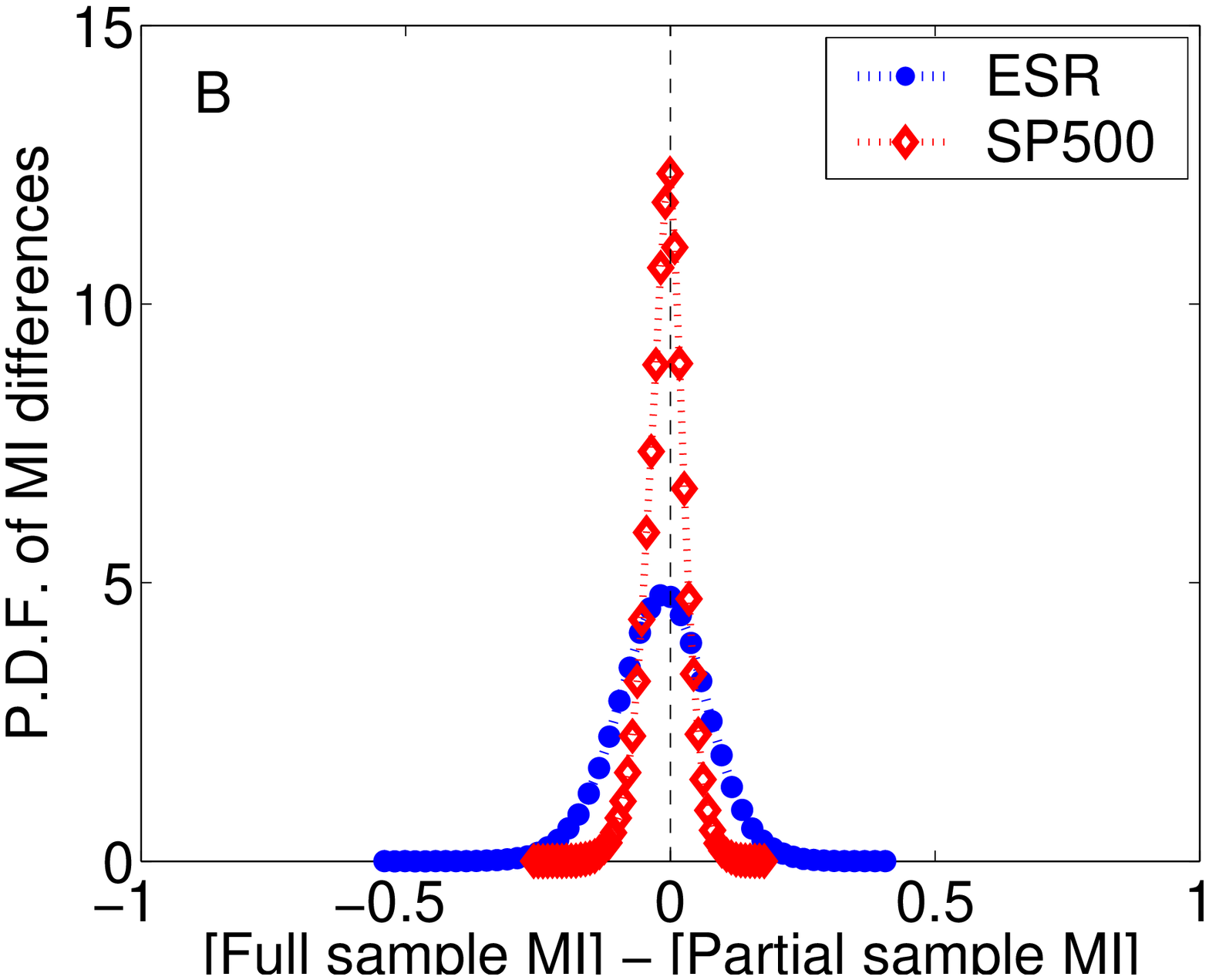,width=.45\columnwidth,height=1.5in}
\end{tabular}
\end{center}
\caption{\small Verification schemes.
{\bf (A)} Probability density (in a logarithmic scale) of $\sim 380,000$ extrapolated MI values obtained
for all expression profiles of yeast ESR genes \cite{Gasch00},
before and after random reshuffling. 
{\bf (B)} Probability density of the differences between the 
MI values obtained for the ESR and the $SP500$ data versus the extrapolated 
values after randomly removing one third of the joint sample for every pair.}
\label{Fig:verification}
\end{figure}

We examine the estimates obtained for the real data
versus those obtained for the same data after random shuffling, as shown in \figref{Fig:verification}A  for the ESR data;
when there are no real correlations the extrapolated MI values
are $\approx 0$. Similar results were obtained for the other datasets. 
More subtly, we  compare the MI values
to those obtained from a smaller fraction of the joint sample  than used in the extrapolation procedure
(\figref{Fig:verification}B). Apparently, using the full joint sample
or using (randomly chosen) two thirds of this sample gives approximately the same results;  e.g. in
the $SP500$ data, 
the estimation differences were greater than $0.1$ bits
for less than $2\%$ of the pairs. 
The ESR results were less stable, probably due to 
the smaller sample size and the fact that Microarray readouts are noisy while reported stock prices are precise.  Nonetheless, even for the ESR data our results seem quite robust.  

\subsection{Sorted MI relations and MI--PC comparison}

\begin{figure}[h] 
\begin{center}
\begin{tabular}{cc}
    \psfig{figure=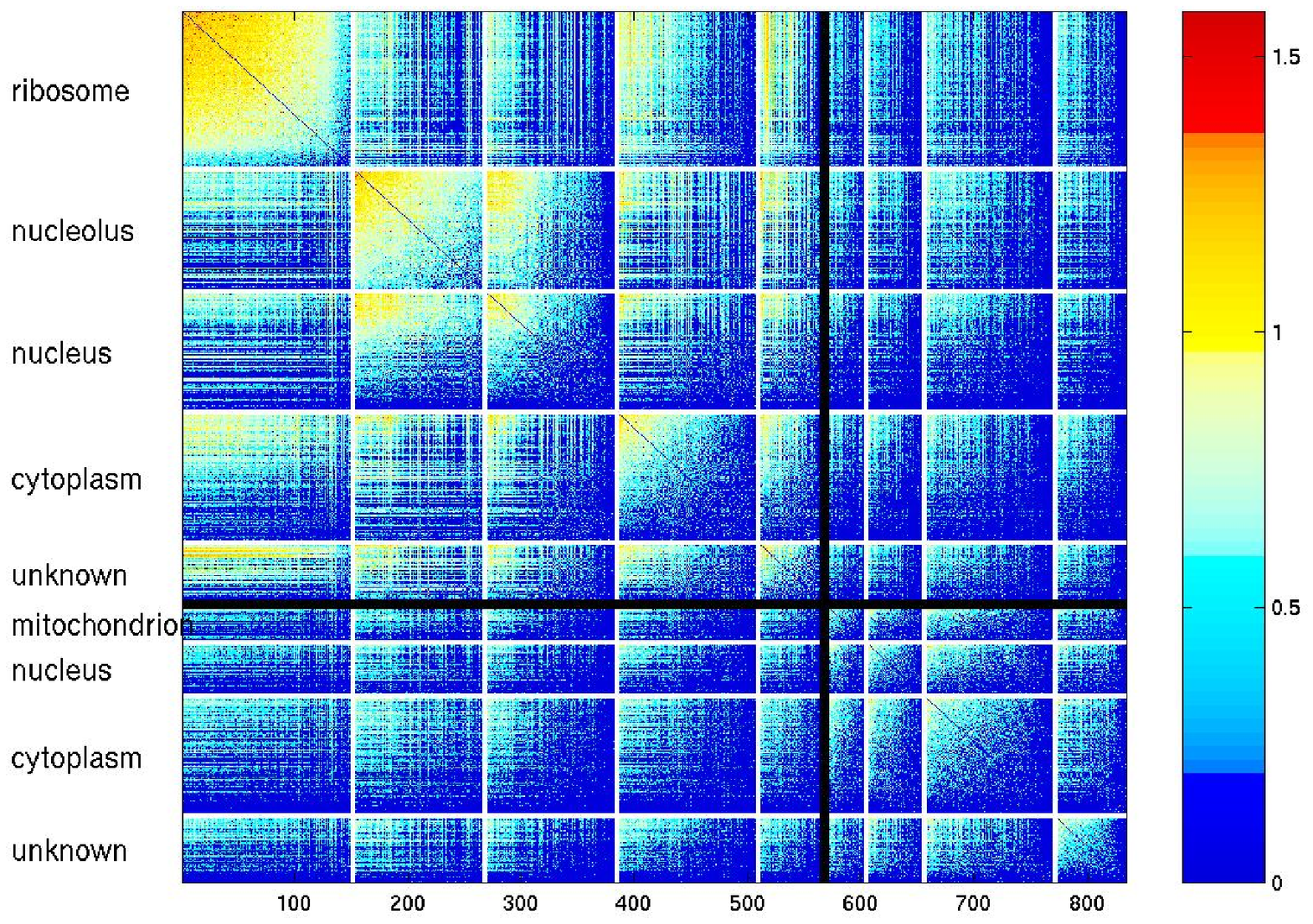,width=.48\columnwidth,height=1.7in} &
    \psfig{figure=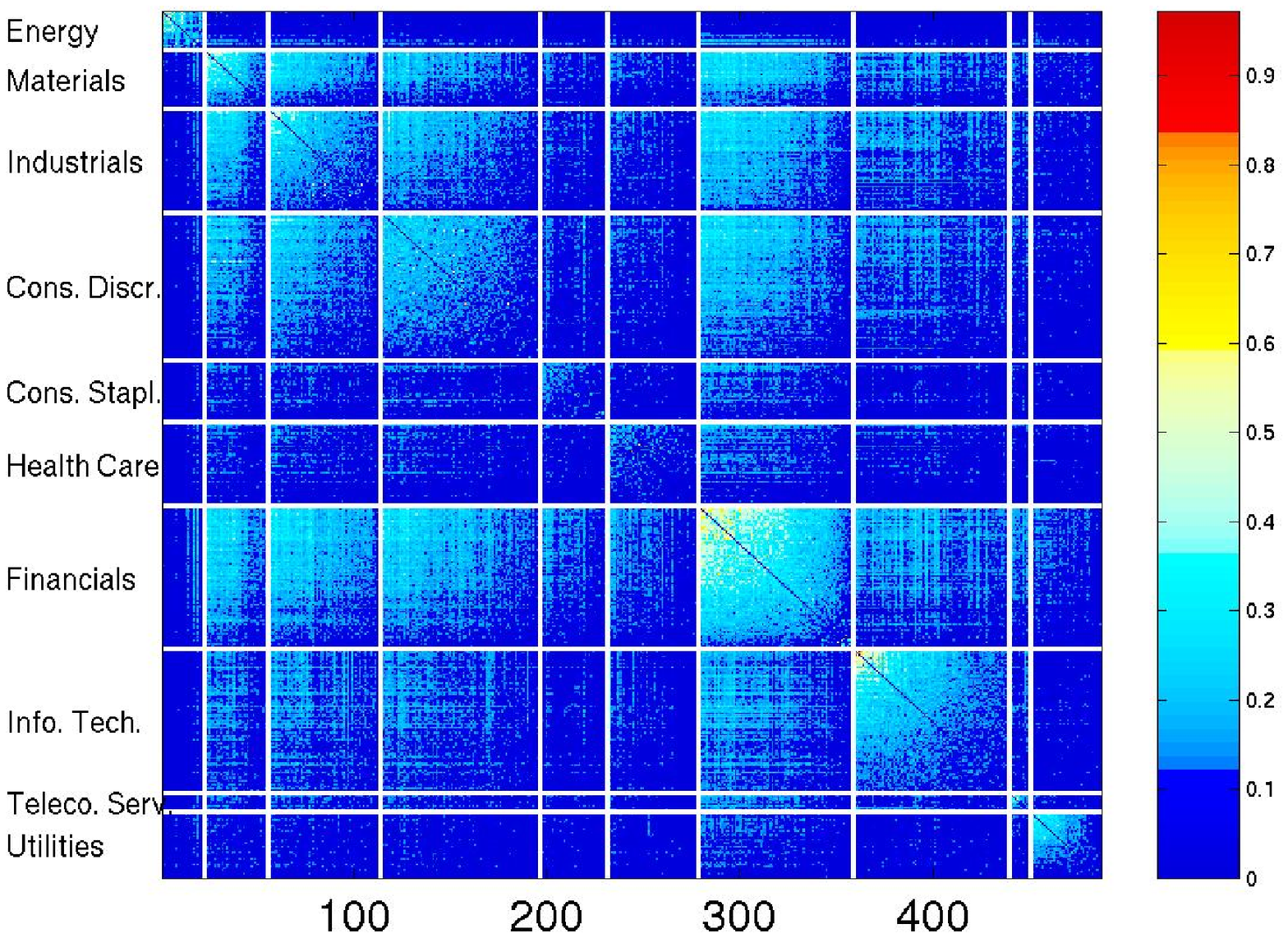,width=.48\columnwidth,height=1.7in}
\end{tabular}
\end{center}
\caption{\small Sorted MI relations. Inside each group, items are 
sorted according to their average MI about other group members.
For brevitiy, MI relations smaller than $\langle I(Y_i;Y_j) \rangle$ are set to zero.
{\bf Left:} ESR MI relations, sorted 
according to (manually chosen) GO cellular-component annotations. The Repressed and the 
Induced modules were sorted independently and are separated by the black solid line. 
{\bf Right:} $SP500$ MI relations sorted according to the 
Standard and Poor's classification.}
\label{Fig:Sorted_I_Relations}
\end{figure}

It is important to ask if patterns of mutual information are meaningful with respect to some external reference.  
When we sort genes by the ``cellular component''  assigned to each gene in the Gene Ontology  \cite{GO}, the matrix of mutual informations in the ESR module acquires a block structure,
indicating that genes that belong to the same cellular component tend to be 
highly informative about each other (\figref{Fig:Sorted_I_Relations} left);
the most tightly connected block correspond to the ribosomal genes.
A similar block structure emerges for the $SP500$ data when we sort stocks according to the 
Standard and Poor's classification of the companies (\figref{Fig:Sorted_I_Relations} right);
this structure matches our intuition about the major sectors of the economy, although
some sectors are significantly better connected
than others (e.g., ``Financials'' vs. ``Health Care'').
The ``Energy'' sector seems quite isolated,
consistent with the fact that 
this sector is heavily regulated and operates under special rules and conditions.
In \tabref{Tbl:EachMoviePairs} we present the $10$ most informative pairs
obtained for the EachMovie data. All these pairs nicely correspond to 
our intuitions about the relatedness of their content and intended audience.

\begin{table*}[h]
\caption{\small The $10$ most informative pairs in the EachMovie data. 
Only pairs with $N > 200$ were considered.}
\label{Tbl:EachMoviePairs}
\scriptsize
\begin{center}
\begin{tabular}{|c|l|l|c|} \hline \hline 
{\bf MI} & {\bf First Movie} & {\bf Second Movie} & {\bf Sample Size}\\ \hline\hline
0.89 & Free Willy & Free Willy 2 & 851\\ \hline
0.59 & Three Colors: Red & Three Colors: Blue & 1691\\ \hline
0.56 & Happy Gilmore & Billy Madison & 1141\\ \hline
0.56 & Bio-Dome & Jury Duty & 280\\ \hline
0.54 & Homeward Bound II & All Dogs Go to Heaven 2 & 735\\ \hline
0.54 & Ace Ventura: Pet Detective & Ace Ventura: When Nature Calls & 7939\\ \hline
0.52 & Return of the Jedi & The Empire Strikes Back & 2862\\ \hline
0.51 & The Brady Bunch Movie & A Very Brady Sequel & 301\\ \hline
0.50 & Snow White & Pinocchio & 3076\\ \hline
0.49 & Three Colors: Red & Three Colors: White & 1572\\ \hline\hline
\end{tabular} 
\end{center}
\end{table*} 

\begin{figure}[t] 
\begin{center}
\begin{tabular}{ccc}
    \psfig{figure=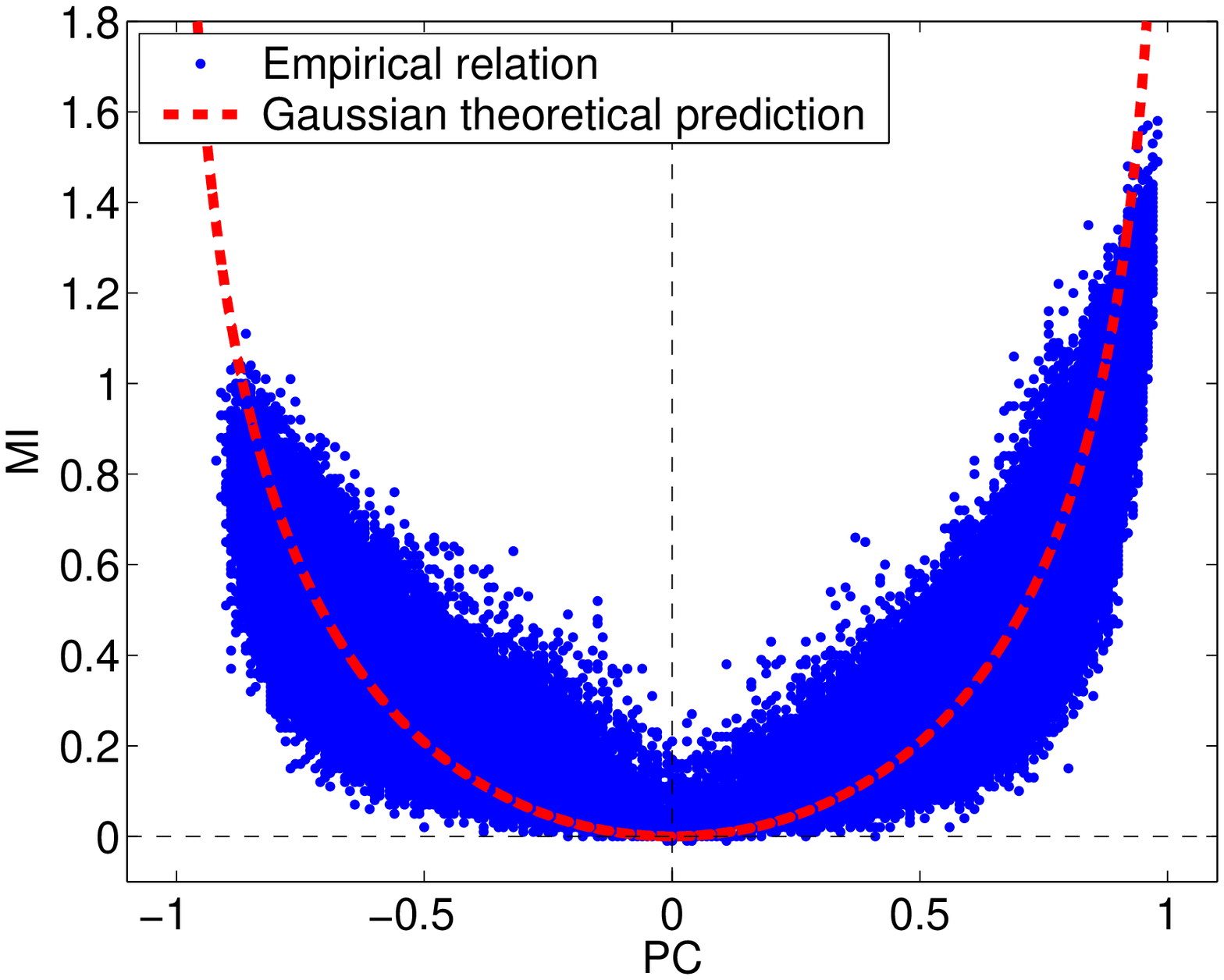,width=.3\columnwidth,height=1.4in} &
    \psfig{figure=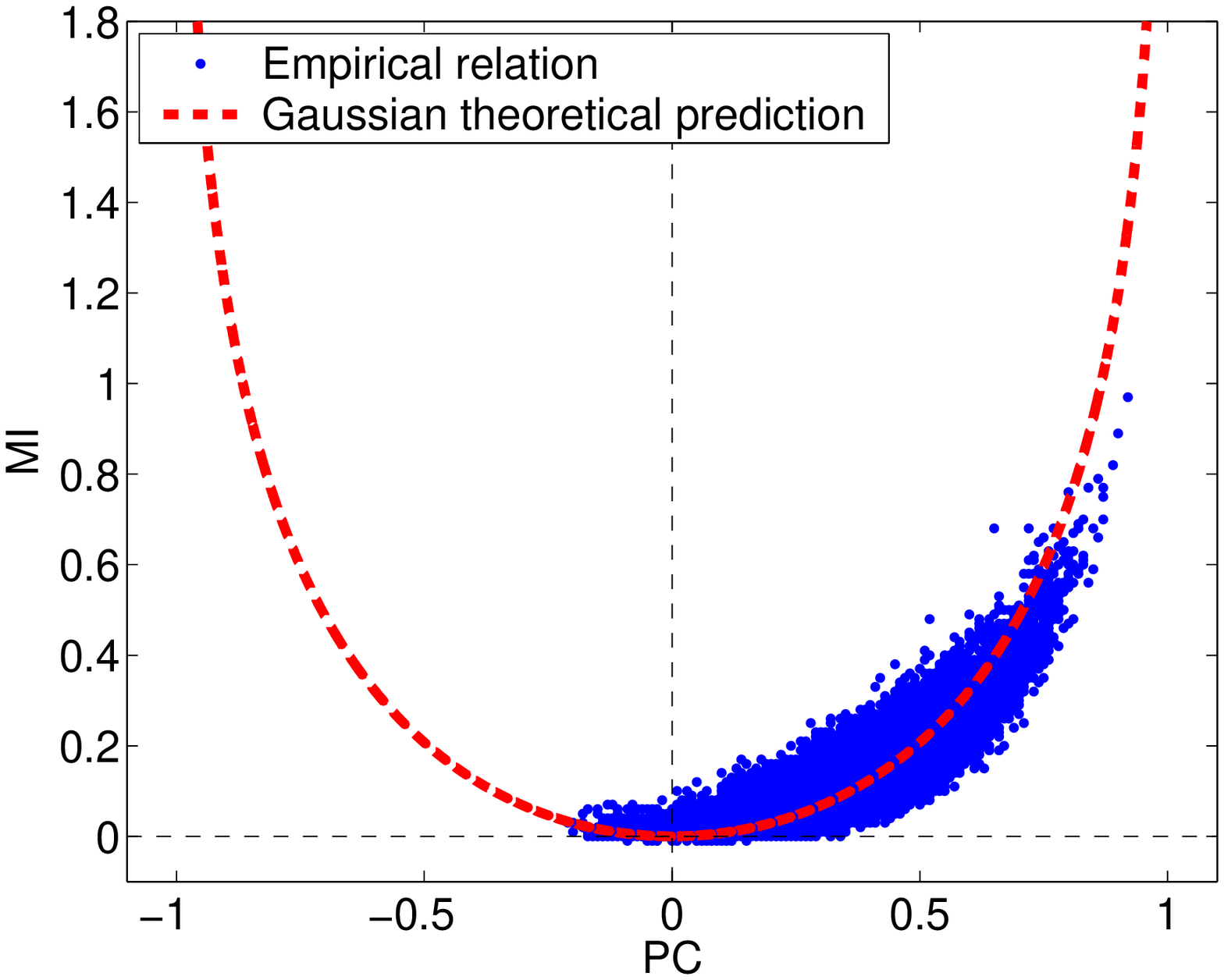,width=.3\columnwidth,height=1.4in} &
    \psfig{figure=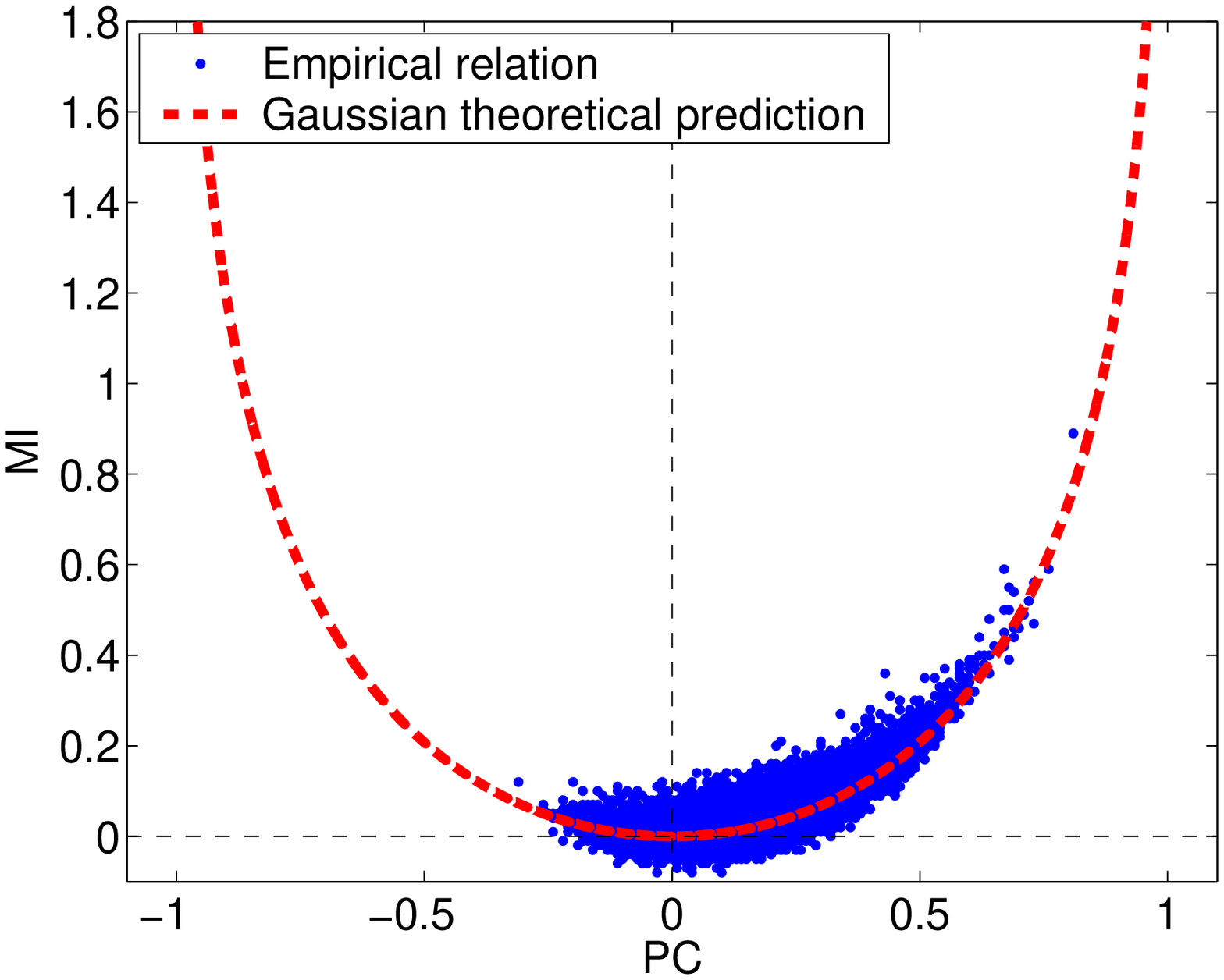,width=.3\columnwidth,height=1.4in} 
\end{tabular}
\end{center}
\caption{\small Comparison of MI and PC relations. {\bf Left:} ESR. {\bf Middle:} $SP500$. 
{\bf Right:} EachMovie (pairs with $N<200$ are not presented).}
\label{Fig:MI_PC}
\end{figure}

It also is interesting to compare the mutual information with a standard correlation measure,
the Pearson Correlation:
$PC(\vec{u},\vec{v}) \equiv {E\left[(u_i-E\left[\vec{u}\right])(v_i-E\left[\vec{v}\right])\right]}/{\sqrt{\Var{\vec{u}}\Var{\vec{v}}}}$; see 
\figref{Fig:MI_PC}.  For Gaussian distributions
we have $I = - ({1}/{2})\log_2(1-PC^2)$ \cite{Kullback68};
this  provides
only a crude approximation of the data,
suggesting that the joint distributions we consider are  
significantly non--Gaussian.  
Note that pairs with relatively large $PC$ and small $I$ are more common
than the opposite, perhaps due to the fact that single outliers suffice
to increase  $PC$ without having a significant effect on  $I$. 
In addition, these results indicate that strong non-linear correlations
(which can be captured only by the MI) were not present in our data.\footnote{At least for the ESR data this is not a surprising result
since the ESR genes are known to be strongly linearly correlated.
In particular, investigating the relations between other genes might yield different results.
An anecdotal example is given in \figref{Fig:EstI}A. Here, the two genes 
(which are not ESR members) have a relatively high MI with a very low PC.}
Finally, we notice that for a given MI (PC) value there is a relatively
large variance in the corresponding PC (MI) values. Thus,
any data analysis  
based on the MI relations is expected to produce different results than PC based analysis. 

\subsection{Results for triplet information}
\label{Sec:TripletI_Results}

In estimating $I_3(y_i,y_j,y_k)$ 
the relevant joint distributions includes $b^3$ parameters;
since $N$ remains the same one must find again an appropriate $b^*$ bound.
In \figref{Fig:EstI}D we present the average triplet information obtained 
for $\sim 10^4$ triplets of randomly reshuffled gene expression profiles \cite{Gasch00}
for different $b^*$ bounds. We used the procedure described in \secref{Sec:TripletEst}
and the same parameter configuration as for pairwise MI estimation. 
The faster growth of this curve as opposed to the same curve for the pairwise relations
demonstrates the ``order of magnitude extra difficulty'' in estimating $I_3$. 
Nonetheless,  $b^*=4$ provides estimates
that properly converge to zero for random data, and at the risk of underestimating some of the $I_3$ values we use this bound in our further analysis.

\begin{figure}[] 
\begin{center}
\begin{tabular}{cc}
    \psfig{figure=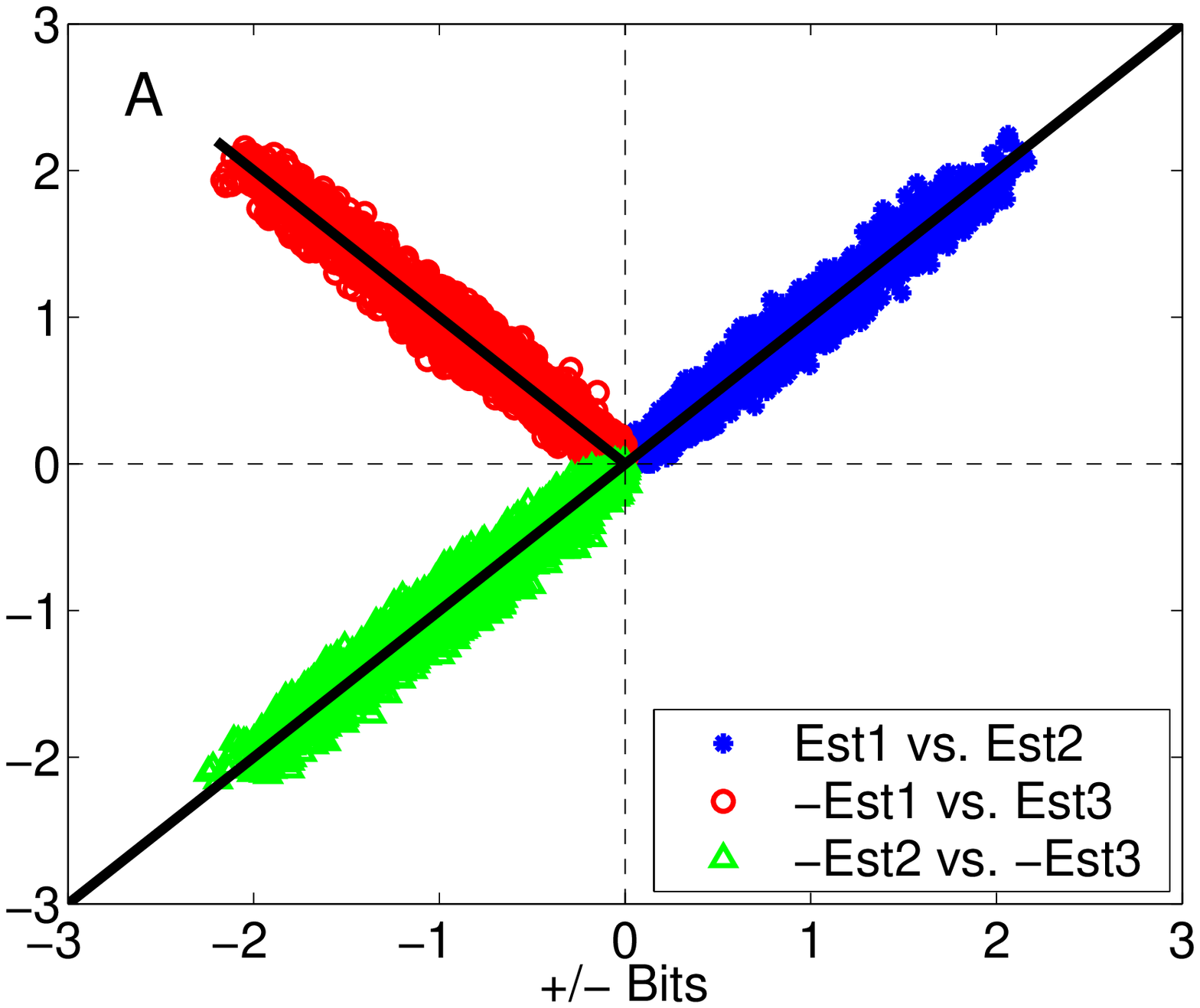,width=.45\columnwidth,height=1.5in} &
    \psfig{figure=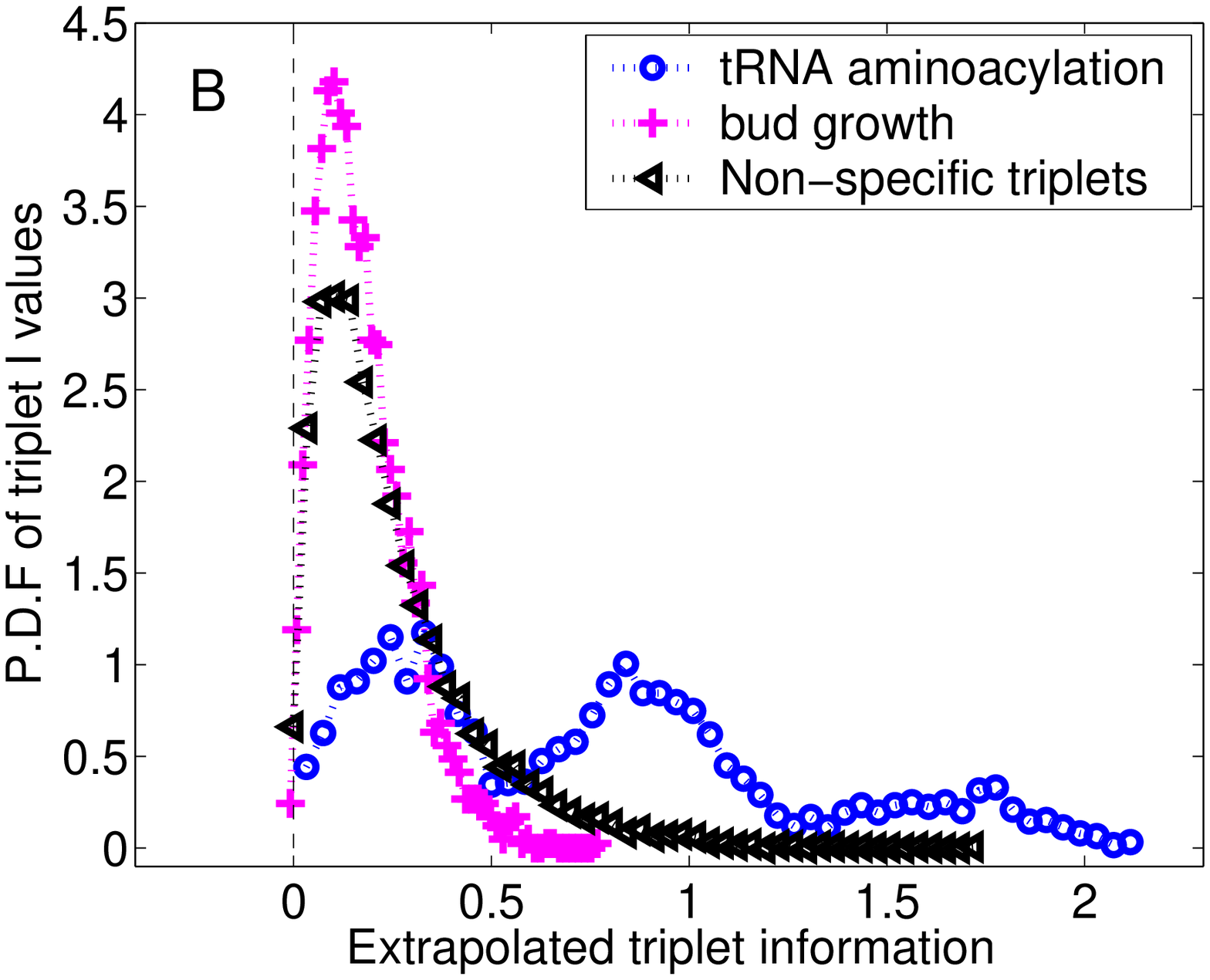,width=.45\columnwidth,height=1.5in} \\
    \psfig{figure=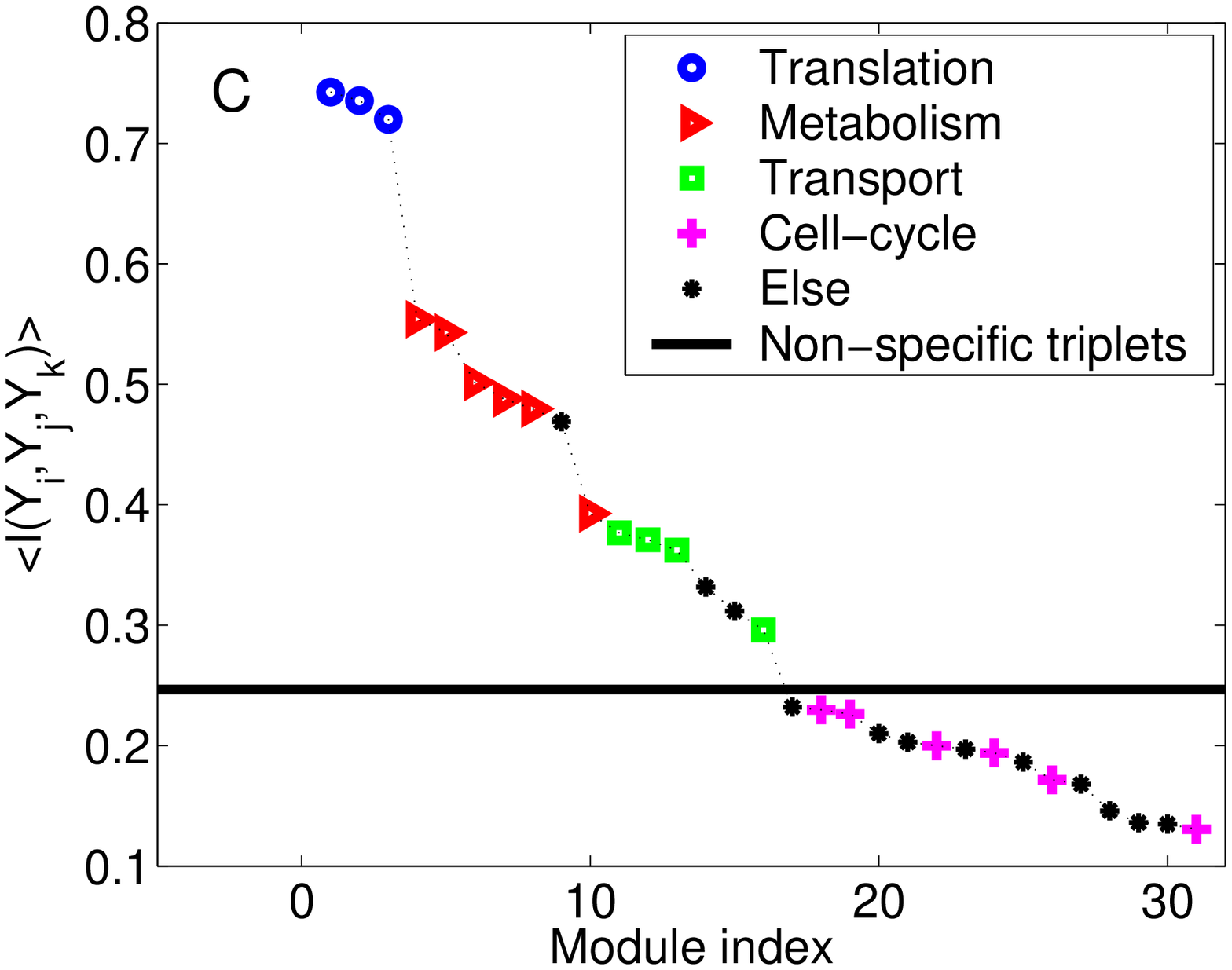,width=.45\columnwidth,height=1.5in} &
    \psfig{figure=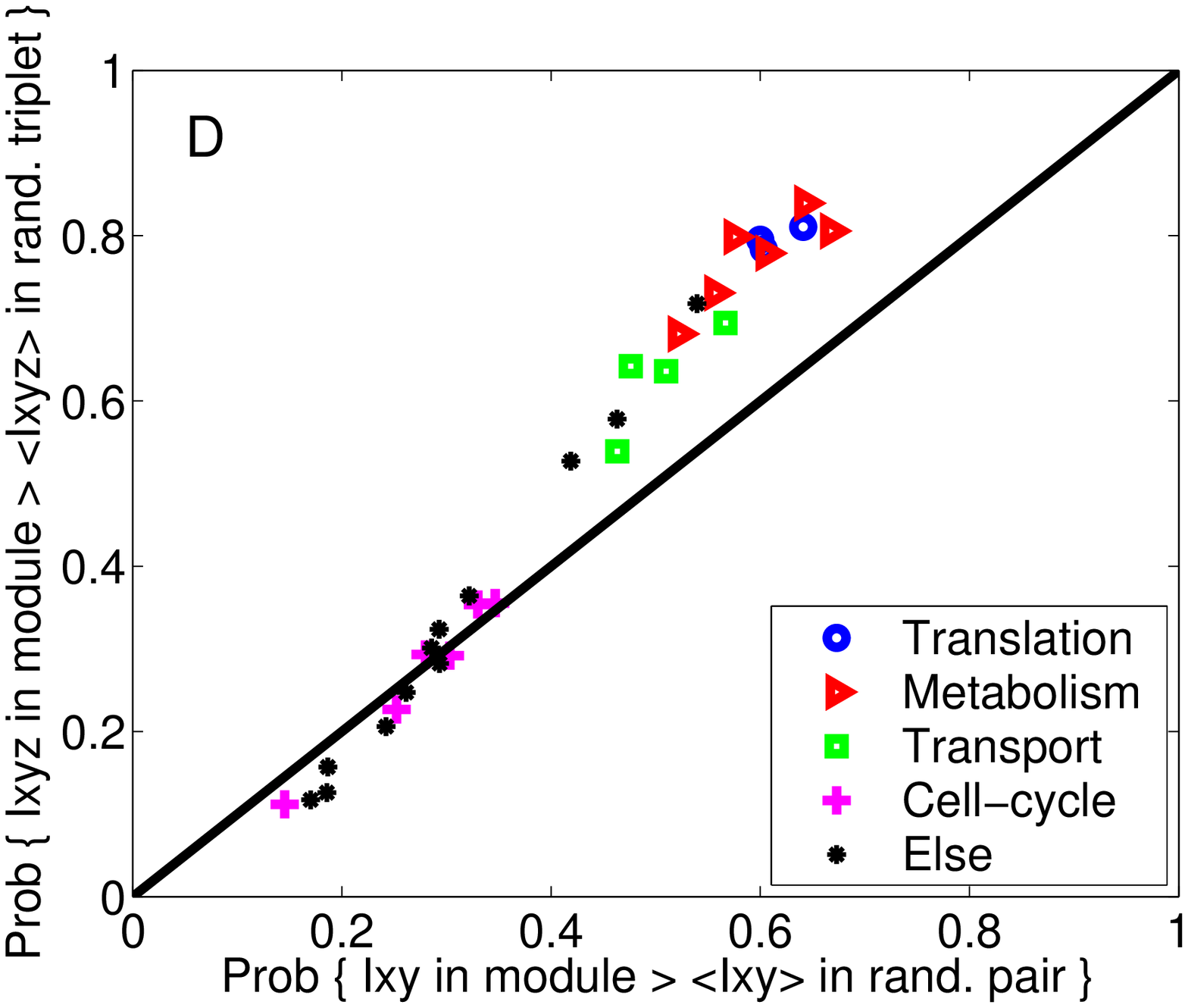,width=.45\columnwidth,height=1.5in} 
\end{tabular}
\end{center}
\caption{\small {\bf (A)} Comparison of the three different estimations of $I_3(y_i,y_j,y_k)$
for the $\approx 3000$ triplets among the $27$ genes annotated 
with 'tRNA aminoacylation for protein translation'.
Similar results were obtained for all other modules.
{\bf (B)} Probability density of extrapolated triplet information values 
for triplets of the `tRNA aminoacylation' module, `bud growth' module and
non--specific triplets.
{\bf (C)} $\langle I_3(y_i,y_j,y_k) \rangle$ values for all modules. 
For modules that highly overlap, only one module is represented 
in the figure. 
The black line represents $\langle I_3(y_i,y_j,y_k) \rangle$ in $\sim 10^4$ non-specific triplets.
{\bf (D)} Vertical axis: Probability of a triplet information value (in a given module) 
to be greater than  $\langle I_3(y_i,y_j,y_k) \rangle$ 
in non-specific triplets. Horizontal axis: Probability 
of a pairwise information value in a given module to be greater than
$\langle I(y_i;y_j) \rangle$ in non-specific pairs.}
\label{Fig:TripletI_GO}
\end{figure}

Computing  all triplet information relations in a given data set might be too demanding, but
computing all the triplet relations in specified subsets is  feasible.
As a test case, we chose all the GO biological process annotations
that correspond to a relatively small set of genes from the entire genome. Specifically, 
$44$ annotations were assigned to $n$ genes with $25 \leq n \leq 30$. 
In each of these $44$ groups we estimated all the 
$I_3$ values,  a total of $\sim 10^5$ estimated relations. 
Recall that every triplet information can be estimated in $3$ different ways
via $3$ different compositions of MI terms;
these three estimations provide consistent results (\figref{Fig:TripletI_GO}A), 
which further support the validity of our procedure, and
we use the average of these $3$ estimates in our analysis.

The distribution of $I_3$ values was quite different for different groups of genes
(\figref{Fig:TripletI_GO}B).
`Bud growth' triplets  
display information values which are even lower
than non--specific triplets (chosen at random from the whole genome), suggesting that most of the bud growth genes
do not act as a correlated module in stress conditions.
For the `tRNA aminoacylation' group  
we see three different behaviors,
suggesting that a subset of these genes correspond to the same regulatory signal.\footnote{Specifically, in such a scenario one should expect to find high
information values for triplets comprised solely of genes from this co--regulated subset,
medium information values for triplets in which only a pair came from this subset,
and low information values for the rest of the triplets.}

In \figref{Fig:TripletI_GO}C we present the average $I_3(y_i,y_j,y_k)$ values
for each group of genes. Interestingly, these values correspond to four relatively
distinct groups. The first, with the highest average information,
comprised of three ``Translation related'' annotations
(like `tRNA aminoacylation').
The second group mainly consisted of  ``Metabolism/Catabolism related'' annotations
(e.g., `alcohol catabolism').
In the third group we find several ``Transport/Export related'' annotations
(e.g., `anion transport'). Finally, in the last group, with the lowest information
values, we have several ``cell-cycle related'' modules (like `bud growth').
These results merit further investigation which will be done elsewhere.

Multi--information can be decomposed into contributions from interactions at different orders, so that, for example,  high $I_3$ can arise due to  high pairwise information relations, but also in situations where there is no information at the pair level \cite{Schneidman+al03}.
In \figref{Fig:TripletI_GO}D we compare the levels of pairwise and triplet information, measuring the probability that pairs or triplets from each group have higher information than randomly chosen,  non--specific pairs or triplets.
Evidently, the gap between these two measures increases monotonically 
as the group becomes more strongly connected, suggesting that a significant portion
of the high triplet information values 
cannot be attributed solely to high pairwise information relations alone.  
Analysis of triplet information relations in the 
$SP500$ and the $EachMovie$ data will be presented elsewhere.

\section{Discussion}
\label{Sec:Discussion}

In principle, mutual and multi--information have several important advantages.  
Information is a domain independent measure which is sensitive to any
type of dependence, including nonlinear relations.  
Information is relatively insensitive to outliers in the measurement space,
and is completely invariant to any invertible changes of variables
(such as the log transformation). Information also  is measured on a physically
meaningful scale:  more than one  bit of information
between two gene expression profiles (\figref{Fig:verification}A) implies
that co--regulation of these genes must involve something
more complex than just turning expression on and off.

The main obstacle is obtaining reliable measurements of these quantities,
especially if there are a lot of relations to consider. 
This paper establishes the use of the direct estimation method in these situations.   More sophisticated estimation tools are available (e.g. \cite{nemenman+al_02}) which allow reliable inference from smaller data sets, but these tools need to be scaled for application to large networks.  
Finally, an important aspect of the work reported here is the estimation of multi--information; we have done this explicitly for triplets, but Eq (\ref{Eq:multiIbreak}) shows us that given sufficient samples the ideas presented here are applicable to all orders.  These collective measures of dependence---and the related concepts of synergy and connected information \cite{Schneidman+al03}---are likely to become even more important as we look at interactions and dynamics in large networks.  

\section*{Acknowledgments}
We thank R Zemel for helpful discussions.  
This work was supported by NIH grant P50 GM071508.
G Tka\v{c}ik acknowledges the support of
the Burroughs-Wellcome Graduate Training Program in Biological Dynamics.

\end{document}